\documentclass[letterpaper, 10 pt, conference]{ieeeconf}  

\IEEEoverridecommandlockouts                              
\usepackage{graphics} 
\usepackage{epsfig} 
\usepackage{mathptmx} 
\usepackage{times} 
\usepackage{amsmath} 
\usepackage{amssymb}  

\usepackage[colorlinks=true, linkcolor=black, urlcolor=blue, citecolor=black]{hyperref}
\usepackage{cleveref}
\usepackage{xcolor}
\usepackage{tikz}
\usepackage{pgfplots}

\usepackage{placeins}    
\usepackage{comment}

\newtheorem{theorem}{Theorem}[section]

\newtheorem{lemma}[theorem]{Lemma}

\newtheorem{remark}[theorem]{Remark}

\newtheorem{propositionA}{Proposition}[section]

\newenvironment{customproof}[1][Proof]
{\par\noindent\textit{#1}\ignorespaces}
{\hfill$\blacksquare$\par}

\pgfplotsset{compat=1.18}
\begin{document}
	
\title{\bf Input Delay Compensation for a Class of Switched Linear Systems via Averaging Exact Predictor Feedbacks$^*$}
	
\author{
Andreas Katsanikakis$^{1}$ and 
Nikolaos Bekiaris-Liberis$^1$
\thanks{$^*$Funded by the European Union (ERC, C-NORA, 101088147). Views and opinions 
expressed are however those of the authors only and do not necessarily reflect those of the 
European Union or the European Research Council Executive Agency. Neither the European 
Union nor the granting authority can be held responsible for them.}
\thanks{$^1$The authors are with the Department of Electrical and Computer Engineering, 
Technical University of Crete, Chania, 73100, Greece. Emails: akatsanikakis@tuc.gr and nlimperis@tuc.gr.}
}
	
\maketitle
	
\begin{abstract}
The key challenges in design of predictor-based control laws for switched systems with arbitrary switching and long input delay are the potential unavailability of the future values of the switching signal (at current time) and the fact that dwell time may be arbitrary. In the present paper, we resolve these challenges developing a new predictor-based control law that is, essentially, an average of exact predictor feedbacks, each one corresponding to an exact predictor-feedback law for a system that operates only in a single mode. Because the predictor state in our control design does not correspond to an exact predictor, stability can be guaranteed under a restriction on the differences among the system’s matrices and controller’s gains. This is an unavoidable limitation, for a switching signal whose future values may be unavailable, when no constraint is imposed on the values of delay and dwell time (as it is the case here). We establish (uniform) stability of the closed-loop system employing a Lyapunov functional. The key step in the stability proof is constructive derivation of an estimate of the mismatch between an exact predictor feedback and the average of predictor feedbacks constructed. We illustrate the performance of the proposed predictor-based control law in simulation, including comparisons with alternative, predictor-based control laws.
\end{abstract}
	
\section{Introduction} \label{sec:intro}

\setcounter{footnote}{1}

Control of systems with input delays and switched dynamics constitutes a critical challenge that is relevant to various engineering applications, such as, for example, robotics \cite{Robot}, power systems \cite{Luo2008}, automotive systems \cite{traffic_flow}, \cite{IOANNOY}, \cite{ACC}, \cite{AICC}, \cite{Obs_P}, \cite{ISS_P}, and even epidemics spreading dynamic systems \cite{SIR_D}. In robotic systems, for instance, in control of manipulators, input delays result in serious performance and stability degradation, in cases when switching between different dynamic regimes is necessary. Similarly, switching (between acceleration/braking modes) and input delay (due to, e.g., engine dynamics) affect simultaneously the dynamics of autonomous vehicles. For this reason, compensation strategies are required in order to ensure stability and efficient operation of systems, under simultaneous presence of switched dynamics and input delay. 

Stability analysis of switched systems with input delays is studied employing Lyapunov-Krasovskii functionals in, for example, \cite{Yue_Kras}, \cite{Mazenc_LKF}, and utilizing Linear Matrix Inequalities (LMIs), in, e.g., \cite{Lin_discrete}, \cite{Xi_Corr}, \cite{Kp}. Control designs that do not rely on employment of infinite-dimensional predictor feedbacks (thus usually restricting the size of input delay) are developed in, for example, \cite{Mazenc},  \cite{Wu_Truncated}, \cite{Sakthi_Truncated}. In \cite{Lin_LTI}, a predictor-feedback control design is proposed, nevertheless relying, for implementation, on the knowledge of the future switching signal values. Our results could be also viewed as related to works dealing with switched, hyperbolic PDE systems; see, for example, \cite{Auriol}, \cite{Prieur}. Perhaps the most closely related, existing results are the ones in \cite{Kong}, \cite{Kong_P1}, \cite{A}, which develop predictor-based control laws for linear systems (without switchings) with switched (stochastic or deterministic) input delay.

The predictor-based control design that we develop here is motivated by the respective design in \cite{Kong} (Chapter 3) for the case of linear systems (without switchings) with switched input delay, which, in general, appears (as reported in \cite{Kong}) to perform more efficiently as compared with a predictor-based law that relies on a predictor state for an average system (i.e., when one first constructs an average system, subsequently constructing a predictor-feedback law for this expected system). Thus, in comparison with \cite{my_ACC}, in which we present a predictor-based control law that relies on employment of a predictor state corresponding to an average system, here we develop a new control design that employs the average of predictor-feedback laws, each one corresponding to the exact predictor feedback for each subsystem (i.e., for a system operating only in a single mode). Since the control design developed in the present paper is different than the one in \cite{my_ACC}, here we introduce a different backstepping transformation (in delay systems representation), we derive new estimates on the bound of the predictor-feedback laws mismatch, between the (inapplicable) exact predictor-feedback law and the average of predictor-feedback laws corresponding to each subsystem, and we implement in simulation a different control law whose performance is also compared with the control design from \cite{my_ACC}.  

More specifically, in the present paper, we consider linear switched systems with long input delay that feature an arbitrary dwell time. Because an exact predictor-feedback law requires employment of the future switching signal values, which may not be available (at current time), we construct a delay-compensating predictor-based control design that is, essentially, an average of exact predictor-feedback laws corresponding to the exact predictor feedbacks for each subsystem. Since our feedback law is not an exact predictor-feedback law, to establish (uniform) stability of the closed-loop system we derive, in a constructive manner, an estimate of the error between our control design and an exact predictor-feedback law. As the dwell time and delay values are not restricted, we have to necessarily impose a restriction on the magnitude of the differences among the system’s matrices and the controller’s gains, to guarantee closed-loop stability. We show however that there is a trade-off, between the allowable differences among system’s matrices and controller’s gains, and the values of delay and dwell time, which we explicitly quantify. We perform simulation studies implementing the predictor-based control design, including comparisons with the predictor-feedback laws constructed assuming that the system operates only in a single mode, as well as a comparison with the respective control design from \cite{my_ACC}. 

We start in Section \ref{sec2} presenting the class of switched systems with input delay considered, together with the predictor-based control design and the respective stability properties of the closed-loop system, which are proven. In Section \ref{sec4} we provide consistent simulation results, including comparisons with alternative, predictor-based control designs. Finally, in Section \ref{sec5} we provide concluding remarks and discuss potential topics of future research.

\section{Control Design and Stability Analysis} \label{sec2}
\subsection{Switched Linear Systems with Input Delay}
In this work, we consider a linear switched system subject to a constant delay in the control input, described by the following dynamics
\begin{equation}\label{1.1}
\dot{X}(t) = A_{\sigma(t)} X(t) + B_{\sigma(t)} U(t-D), 
\end{equation}
where $X(t) \in \mathbb{R}^n$ is the state vector, $U \in \mathbb{R}$ is the control input, and $D>0$ is arbitrarily long input delay. The switching signal $\sigma:[0,+\infty)  \rightarrow L= \{0,1,2...,l\}$, is a right-continuous piecewise constant function indicating which subsystem is active at time $t$. We assume that no jump occurs in the state at a switching time and that the switching signal follows an arbitrary dwell time condition with $\tau_d > 0$, ensuring that the system switches no more than finitely many times within any finite-time interval \cite{Liberzon}.\footnote{There is no conceptual obstacle to consider switching signals with an average dwell time, as we impose no restriction on the value itself of the dwell time. However, for simplicity, we assume an arbitrary dwell time.} The challenge in this setup is that at time $t$, the future signal $\sigma(t+s)$, where $s>0$, is unknown.
\subsection{Predictor-Based Design via Averaging Predictor Feedbacks}
The system (\ref{1.1}) operates under the following proposed predictor-based controller 
\begin{equation}\label{1.5}
U(t) = \sum_{i=0}^l \frac{K_i\hat{P_i}(t)}{(l+1)},
\end{equation}
where $K_i$ is a selected feedback gain chosen as for $A_i + B_i K_i$ to be Hurwitz and $\hat{P_i}(t)$ is the predictor state, for each subsystem, given by
\begin{equation}\label{subprediction}
\hat{P_i}(t) = e^{A_i D} X(t) + \int_{t-D}^{t} {e^{A_i(t-s)}B_i U(\theta) \, d\theta}.
\end{equation}
 The design is thus based on calculating the exact predictor for each subsystem $i$ and then averaging over all these predictors. The motivation for design (\ref{1.5}) comes from the fact that it provides a simple control law that incorporates a delay-compensating mechanism, relying on averaging the exact predictors in the case in which the system would operate always in a single mode. We note that the exact predictor state is not implementable as it would require availability of the future switching signal values. However, as (\ref{1.5}) does not correspond to an exact predictor, one has to impose certain limitations on the system's matrices and controller's gains. 
 
\subsection{Stability Statement and Proof}
\begin{theorem}\label{Exponential Stability}
Consider the closed-loop system (\ref{1.1}) with the controller (\ref{1.5}), (\ref{subprediction}). Let the pairs $\left(A_i,B_i\right)$ be controllable and choose $K_i$ such that $A_i + B_i K_i$ are Hurwitz, for $i=0,\ldots,l,$ and such that there exist common $ P = P^T > 0 $, $ Q = Q^T > 0 $, satisfying 
    \begin{equation}\label{mean_delayf_stability}
        \left(A_i + B_i K_i \right)^T P + P \left(A_i + B_i K_i\right) \leq -Q.
    \end{equation}
    There exists $\epsilon^* > 0$ such that for any $\epsilon < \epsilon^*$, where
\begin{equation}\label{eps} 
    \epsilon= \max \limits_{i,j=0, \ldots, l} \{ |A_i-A_j|, |B_i-B_j|, |K_i-K_j|\},
\end{equation} and for all $X_0\in\mathbb{R}^n$, $U_0\in L^2[-D,0]$ there exist positive constants $\rho,\xi$ such that the following holds
\begin{align}\label{stability equation}
\left| X(t) \right| + \sqrt{\int_{t-D}^{t} U(\theta)^2 d\theta} &\leq \rho \left( \left| X(0) \right| + \sqrt{\int_{-D}^{0} U(\theta)^2 d\theta} \right) \notag \\& \qquad \times e^{-\xi t}, \quad t \geq 0. 
\end{align}

\end{theorem}
\begin{remark}
     Despite the restriction on $\epsilon$ in (\ref{eps}), no constraints are imposed in Theorem \ref{Exponential Stability} regarding the delay value or dwell time (even though a trade-off exists between the allowable values of $\epsilon$ and the values of delay and dwell time; see Lemma \ref{trans_stability}). Such a restriction is unavoidable, at least without imposing a restriction on dwell time, which is explained as follows. The restriction in Theorem \ref{Exponential Stability} originates in the facts that the switching signal is arbitrary and its future values unavailable. For these reasons the restriction on $\epsilon$ is imposed in order for (\ref{1.5}) to be close to the exact, predictor-feedback law and in order for the choice of control gains to not be far from the actual stabilizing gains of each mode. One could potentially choose a single gain, in order to not restrict the gains differences (and to also obtain a simpler formula for the controller). Nevertheless, in such a case, the differences of the $A_i$ and $B_i$ matrices would have to be restricted further.
\end{remark}

The proof of Theorem \ref{Exponential Stability} relies on some lemmas, which are presented next, together with their proofs. Note that Lemma 2.3 below (and its proof) can be found in \cite{my_ACC} (Lemma 3.3). For the reader’s convenience however we recall it here as well.
\begin{lemma}[{exact predictor construction}]\label{lemma exact predic}
Let the system (\ref{1.1}) experience $k$-switches within the interval $[t,t+D)$, $k \in \mathbb{N}_0$. Then the exact predictor $P(t)$ of this system is
\begin{align}\label{P(t)}
    P(t) &= \prod_{n=1}^{k+1} e^{A_{m_n}(s_n-s_{n-1})}X(t) \notag \\
         &\quad + \sum_{n=1}^{k+1} \left( \prod_{j=n}^{k} e^{A_{m_{j+1}}(s_{j+1}-s_{j})} 
         \int_{t-D+s_{n-1}}^{t-D+s_n} e^{A_{m_n}(t-D+s_n-\theta)} \right. \notag \\
         &\qquad \left. \times B_{m_n} U(\theta) d\theta \right),
\end{align}
where $m_i \in L$, for $i=1,2,\ldots,k+1$, denotes the mode of the system before the $i$-th switching, and $s_i \in \mathbb{R}$, for $i=1,2,\ldots,k$, denotes the $i$-th switching instant, with $s_0=0$ and $s_{k+1}=D$.

\begin{proof}
    As shown in Fig. \ref{fig1}, setting $t=\theta + D$ and $P(\theta)=X(\theta+D)$, then system (\ref{1.1}) becomes
    
    \begin{equation}\label{dP_theta}
        \frac{dP(\theta)}{d\theta}=A_{\sigma_{(\theta+D)}} P(\theta) + B_{\sigma_{(\theta+D)}} U(\theta).
    \end{equation}
    We divide the interval $[t, t+D]$ in intervals of constant modes, such that the system operates in $m_i$ mode if
    \begin{equation}\label{2.5}
       t-D+s_{i-1} \leq \theta < t-D+s_{i}, \quad i=1,2,\ldots,k+1.
    \end{equation}
    In each sub-interval the system does not exhibit switching, and thus, we set ${m_i}=\sigma(\theta+D),$  for $i=1,...,k+1$. We can now proceed to the solution of (\ref{dP_theta}) for each subsystem that is extracted from the standard form of the general solution for a time-invariant ODE system as 
    \begin{align}\label{P_theta}
        P(\theta) &= e^{{A_{m_i}}(\theta-t+D-s_{i-1})} X(t+s_{i-1}) \notag \\
        &\quad + \int_{t-D+s_{i-1}}^{\theta} e^{{A_{m_i}}(\theta - s)} {B_{m_i}}U(s) d s.
    \end{align}
    Setting $i=k+1$ in (\ref{2.5}) and $\theta=t$, from (\ref{P_theta}) we reach 
    \begin{align}\label{Pt_small}
        P(t)&=e^{A_{k+1}(D-s_{k})} X(t+s_{k}) \notag \\ &\quad +\int_{t-D-s_k}^{t} e^{A_{k+1}(t-\theta)} B_{k+1} U(\theta) d \theta .
    \end{align}     
\begin{figure}[t]
    \centering
    \includegraphics[width=8.5 cm]{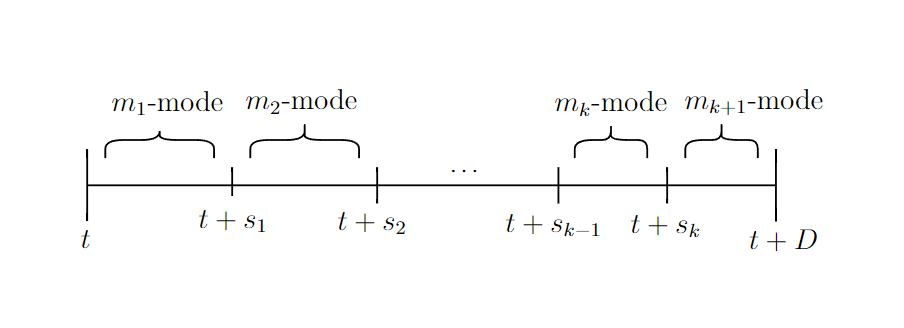}
    \caption{Switching instants and respective modes in the $[t, t+D]$ time interval.
    \label{fig1}}
\end{figure}
Additionally, for $\theta = t-D+s_i$ from (\ref{P_theta}), we arrive at
\begin{align}\label{2.8}
    X(t+s_i) &= e^{A_{m_i}(s_i-s_{i-1})} X(t+s_{i-1}) \notag \\
    &\quad + \int_{t-D+s_{i-1}}^{t-D+s_i} e^{A_{m_i}(t-D + s_i - \theta)} B_{m_i} U(\theta) d\theta.
\end{align}
    Eventually (\ref{2.8}) can be written in terms of $X(t)$ as
    \begin{align} \label{X(t+si)}
        X(t+s_i) &= \prod_{n=1}^{i}{e^{{A_{m_n}}(s_n-s_{n-1})}} X(t) \notag \\
                 &\quad + \sum_{n=1}^{i}\prod_{j=n}^{i-1}{e^{{A_{m_{j+1}}}(s_{j+1}-s_{j})}}\int_{t-D+s_{n-1}}^{t-D+s_n}e^{A_{m_n}(t-D+s_n-\theta)} \notag \\ &\qquad \times B_{m_n}U(\theta)d\theta,
    \end{align}
    for $t-D+s_{i-1} \leq \theta \leq t-D+s_i$, where $i=1,...,k+1$,  $s_0=0$ and  $s_{k+1}=D$. Setting $i=k+1$, we have $X(t+s_{k+1})=X(t+D)=P(t)$. Hence from (\ref{X(t+si)}) we arrive at (\ref{P(t)}).    
\end{proof}
\end{lemma}

\begin{lemma}[{backstepping transformation}] 
The following backstepping transformation, 
    \begin{equation}\label{W_theta}
        W(\theta)=U(\theta) - K_{\sigma{(\theta+D)}} P(\theta), \quad t-D \leq \theta \leq t,
    \end{equation}
    where $P(\theta)$ is obtained from (\ref{P_theta}) for each $\theta$ interval defined in (\ref{2.5}),  together with the control law (\ref{1.5}), transforms system (\ref{1.1}) to the target system 
    \begin{align}
        \dot{X}(t) &= \left(A_{\sigma(t)}+B_{\sigma(t)}K_{\sigma{(t)}}\right) X(t) + B_{\sigma(t)} W(t-D) \label{Xd_trans} , \\
              W(t) &= U(t)-K_{\sigma{(t+D)}}P(t) ,\quad t\geq0, \label{W_trans}
    \end{align}
where $U(t)$ and $P(t)$ are given in (\ref{1.5}) and (\ref{P(t)}), respectively.

\begin{proof}
    System (\ref{1.1}) can be written as
    \begin{align}
        \dot{X}(t) &= \left(A_{\sigma(t)} + B_{\sigma(t)}K_{\sigma{(t)}}\right) X(t) \notag \\
                   &\quad + B_{\sigma(t)} \left(U(t-D) -  K_{\sigma{(t)}}X(t)\right). \label{2.16}
    \end{align}
    We now use (\ref{W_theta}). Setting $\theta=t-D$, from (\ref{P_theta}) we get $P(t-D)=X(t)$. Observing (\ref{P_theta}) and (\ref{2.16}), transformation (\ref{W_theta}) maps the closed-loop system consisting of the plant (\ref{1.1}) and the control law (\ref{1.5}), to the target system (\ref{Xd_trans}), (\ref{W_trans}).
\end{proof}
\end{lemma}

\begin{lemma}[{inverse backstepping transformation}]\label{inverse_transformation} 
The inverse backstepping transformation of $W$ is
\begin{equation}\label{inverse_theta}
    U(\theta)=W(\theta) + K_{\sigma{(\theta+D)}} \Pi(\theta),
\end{equation}
where for $t-D \leq \theta \leq t$ divided in the sub-intervals (\ref{2.5}),
\begin{align}\label{2.19}
    \Pi(\theta) &= e^{({A_{m_i}+B_{m_i}K_{m_i})}(\theta-t+D-s_{i-1})} X(t+s_{i-1}) \notag \\
    &\quad +\int_{t-D+s_{i-1}}^{\theta} e^{({A_{m_i}+B_{m_i}K_{m_i})}(\theta - s)} {B_{m_i}}W(s) d s.
\end{align}

\begin{proof}
    Firstly, we observe from (\ref{W_theta}) that $U(\theta)=W(\theta) + K_{\sigma{(\theta+D)}} P(\theta)$. Solving the ODE (\ref{Xd_trans}) in a similar way as (\ref{dP_theta}) it can be shown that $\Pi(\theta)=X(\theta+D)$, where $\Pi(\theta)$ is given from (\ref{2.19}), and it holds that $\Pi(\theta)=P(\theta)$. 
\end{proof}
\end{lemma}

\begin{lemma}[{bound on error due to predictor mismatch}]\label{lemma_w(t)}
Variable $W(t)$ defined in (\ref{W_trans}) satisfies
\begin{equation}\label{Wt_bound}
    | W(t) | \leq \lambda(\epsilon) \left(  |X(t)| +  \int_{t-D}^{t}{|U(\theta)|d\theta} \right), \quad t \geq 0,
\end{equation}
where $\lambda : \mathbb{R}_+ \rightarrow \mathbb{R}_+$ is a class $K_{\infty}$ function and $\epsilon$ is defined in (\ref{eps}).

\begin{proof}
    For $W(t)$ defined in (\ref{W_trans}), using (\ref{P(t)}) we write
    \begin{equation}\label{Wt}
        W(t) = \frac{1}{l+1} \left( \Delta_1(t) + \Delta_2(t) \right),
    \end{equation}
    where 
    \begin{align}
        \Delta_1(t) &= \sum_{i=0}^l{ \left(K_i \prod_{n=1}^{k+1}{e^{{A_i}(s_n-s_{n-1})}}- K_{\sigma{(t+D)}} \prod_{n=1}^{k+1}{e^{{A_{m_n}}(s_n-s_{n-1})}}\right)} \notag \\
        & \quad \times X(t),  \label{D1t1} \\
        \Delta_2(t) &= \sum_{i=0}^l \sum_{n=1}^{k+1} \left( K_i  \prod_{j=n}^{k}{e^{A_i(s_{j+1}-s_{j})}}\int_{t-D+s_{n-1}}^{t-D+s_n}e^{A_i(t-D+s_n-\theta)} \right. \notag \\ 
                    &\qquad \times B_i U(\theta)d\theta  \notag \\ 
                    &\quad - K_{\sigma{(t+D)}} \prod_{j=n}^{k}{e^{{A_{m_{j+1}}}(s_{j+1}-s_{j})}}\int_{t-D+s_{n-1}}^{t-D+s_n}e^{A_{m_n}(t-D+s_n-\theta)} \notag \\
                    &\qquad \left. \times B_{m_n}U(\theta)d\theta\right) \label{D2t1}.
    \end{align}
    Since $m_{k+1}=\sigma(t+D)\in L$, $t \geq 0$, we can re-write (\ref{D1t1}) as
        \begin{align}
        \Delta_1(t) &= \sum_{i=0}^l \left\{ K_i \left( \prod_{n=1}^{k+1}{e^{{A_i}(s_n-s_{n-1})}} -\prod_{n=1}^{k+1}{e^{{A_{m_n}}(s_n-s_{n-1})}}\right) \right. \notag \\
                    &\quad \left. + \Delta K_{i,m_{k+1}}\prod_{n=1}^{k+1}{e^{{A_{m_n}}(s_n-s_{n-1})}} \right\} X(t), \label{D1t}
        \end{align}
    where for any matrix $R_i$ we set
    \begin{align}
        \Delta R_{i,j} &= R_{i}-R_{j}, \label{Rmn}\\
         \epsilon_{R_{_{i,j}}} &= \left| \Delta R_{i,j} \right|, \label{epsR1}  \\
         \epsilon_R &= \max \limits_{i,j=0,\ldots,l} \{\epsilon_{R_{i,j}}\}. \label{epsR}      
    \end{align}    
    Applying the following property in each part of (\ref{D2t1}), 
    \begin{align}
        M'N'-MN=M'(N'- N) + (M'-M)N,
    \end{align}
    where $M,N,M',N'$ denote arbitrary matrices, we write (\ref{D2t1}) as
    \begin{align}\label{d2t_new}
        \Delta_2(t) &= \sum_{i=0}^{l} \sum_{n=1}^{k+1} \left\{ K_i \left\{ \prod_{j=n}^{k} { e^{{A_i}(s_{j+1}-s_{j})}}\left(Z_{1,n}(t) +Z_{2,n}(t)\right) \right. \right. \notag \\ & \quad \left. \left.  + Z_{3,n}(t)\right\} +\Delta_{K_{i,m_{k+1}}} \left(  \prod_{j=n}^{k}{e^{A_i(s_{j+1}-s_{j})}} \right. \right. \notag \\
        &\quad \left. \left.  \times \int_{t-D+s_{n-1}}^{t-D+s_n}e^{A_i(t-D+s_n-\theta)} B_i U(\theta)d\theta \right)  \right\},
    \end{align}
    where
    \begin{align}
        Z_{1,n}(t) &=  \int_{t-D+s_{n-1}}^{t-D+s_n}{e^{A_i(t-D+s_n-\theta)}(B_i-B_{m_n})U(\theta)d\theta},\label{Z1}  \\
        Z_{2,n}(t) &=  \int_{t-D+s_{n-1}}^{t-D+s_n}\left( e^{A_i(t-D+s_n-\theta)} -e^{A_{m_n}(t-D+s_n-\theta)} \right) \notag \\ 
        &\qquad \times B_{m_n}U(\theta)d\theta, \label{Z2}\\
        Z_{3,n}(t) &=  {\left(\prod_{j=n}^{k}{e^{{A_i(s_{j+1}-s_{j})}}}-\prod_{j=n}^{k}{e^{{A_{m_{j+1}}}(s_{j+1}-s_{j})}}\right)} \notag \\ 
        &\qquad \times \int_{t-D+s_{n-1}}^{t-D+s_n}e^{A_{m_n}(t-D+s_n-\theta)}B_{m_n}U(\theta)d\theta  .  \label{Z3}
    \end{align}
    For any matrix $R_i$, where $R$ can be $A,B$ or $K$, define 
    \begin{align}        
            M_R &= \max \{|R_0|,|R_1|,\ldots,|R_l|\}. \label{M_A} 
    \end{align}    
    Setting $Y_1=A_{m_i}(s_i-s_{i-1}), \, Y_2=-\Delta A_{m_i}(s_i-s_{i-1})$, for $m_i$, $s_i$ defined in Lemma \ref{lemma exact predic}, and applying the result from Proposition \ref{prop1} (in the Appendix) we arrive at
    \begin{align}\label{1st_diffs}
        \left |e^{A_i(s_n-s_{n-1})} - e^{A{m_n} (s_n-s_{n-1})} \right | &\leq \epsilon_{A_{i,m_n}} (s_n-s_{n-1}) e^{|A_m{_n}|(s_n-s_{n-1})} \notag \\ 
          &\qquad \times e^{\epsilon_{A_{i,m_n}}(s_n-s_{n-1})}.
    \end{align}
    Recalling that $(s_n-s_{n-1}) \leq D$, and substituting (\ref{epsR}), (\ref{M_A}) in (\ref{1st_diffs}) we get
    \begin{equation}\label{TheorResult}
        \left |e^{A_i(s_n-s_{n-1})} - e^{A{m_n} (s_n-s_{n-1})} \right | \leq \epsilon_{A} \cdot D \cdot e^{ M_A (s_n-s_{n-1}) } e^{ \epsilon_{A} D }.
    \end{equation}  
    We now upper bound the expression from (\ref{D1t}). We define
    \begin{equation}\label{Tk+1}
        T_{k+1}=\left| \prod_{n=1}^{k+1}{e^{{A_i}(s_n-s_{n-1})}}- \prod_{n=1}^{k+1}{e^{{A_{m_n}}(s_n-s_{n-1})}} \right|.
    \end{equation}
    Developing (\ref{Tk+1}) for each iteration we have
    for $k=0$
        \begin{equation}\label{n=1}
            T_1 = \left| e^{A_is_1} - e^{A{m_1} s_1} \right|. 
        \end{equation}
        The result in (\ref{TheorResult}) can be directly applied to (\ref{n=1}). Thus,
        \begin{align}\label{n=1Result}
            T_1 \leq \epsilon_{A} \cdot D \cdot e^{ M_A s_1 } e^{ \epsilon_{A} D }.
        \end{align}
    For $k=1$
        \begin{align}
            T_2 = \left| e^{A_is_1}e^{A_i(s_2-s_1)} - e^{A{m_1}s_1} e^{A{m_2}(s_2-s_1)} \right|.
        \end{align}
        We expand the difference within the norm and using the triangle inequality for the norm bounds we get \begin{align}\label{n=2}
            T_2 & \leq \left| e^{A_is_1} \right| \cdot \left| e^{A_i(s_2-s_1)} - e^{A{m_2} (s_2-s_1)} \right| \notag \\ 
            & \quad + \left| e^{A{m_2} (s_2-s_1)} \right| \cdot \left| e^{A_is_1} - e^{A{m_1} s_1} \right| .
        \end{align}
        We apply now (\ref{TheorResult}), (\ref{n=1Result}) to (\ref{n=2}), to obtain
        \begin{align}
            T_2 \leq \epsilon_{A} \cdot 2D \cdot e^{ M_A s_2 } e^{ \epsilon_{A} D } .
        \end{align}
    For some $k$ we assume that the following expression holds
        \begin{align}\label{tkResult}
            T_{k} \leq \epsilon_{A} \cdot kD \cdot e^{ M_A s_k } e^{ \epsilon_{A} D } .
        \end{align}
    We prove that the formula holds generally using the induction method.     
    Since (\ref{n=1Result}), (\ref{tkResult}) hold, expanding (\ref{Tk+1}) we get
    \begin{align}
        &T_{k+1}  \notag \\
            &= \left| e^{A_i(s_{k+1}-s_k)}\prod_{n=1}^{k}{e^{{A_i}(s_n-s_{n-1})}}- e^{A_{m_{k+1}}(s_{k+1}-s_k)}\prod_{n=1}^{k}{e^{{A_i}(s_n-s_{n-1})}} \right. \notag\\
            &\quad +  e^{A_{m_{k+1}}(s_{k+1}-s_k)}\prod_{n=1}^{k}{e^{{A_i}(s_n-s_{n-1})}} - e^{A_{m_{k+1}}(s_{k+1}-s_k)} \notag \\ 
            &\qquad \left. \times \prod_{n=1}^{k}{e^{{A_{m_n}}(s_n-s_{n-1})}} \right|.
    \end{align}
        The triangle inequality and the submultiplicative properties hold hence,
        \begin{align}\label{tk+1last}
            T_{k+1}&\le \left| e^{A_i(s_{k+1}-s_k)}-e^{A_{m_{k+1}}(s_{k+1}-s_k)}\right|\left|\prod_{n=1}^{k}{e^{{A_i}(s_n-s_{n-1})}}\right| \notag \\ 
                   &\quad + \left|e^{A_{m_{k+1}}(s_{k+1}-s_k)}\right| \left|\prod_{n=1}^{k}{e^{{A_i}(s_n-s_{n-1})}} - \prod_{n=1}^{k}{e^{{A_{m_n}}(s_n-s_{n-1})}} \right|.
        \end{align}
        Applying (\ref{TheorResult}) and (\ref{tkResult}) to (\ref{tk+1last}) we get
        \begin{align}\label{tk+1Result}
            T_{k+1}&\le \left( \epsilon_{A} \cdot D \cdot e^{ M_A (s_{k+1}-s_{k}) } e^{ \epsilon_{A} D } \right) e^{M_A s_k}  + e^{M_A (s_{k+1}-s_k)} \notag \\
            &\qquad \times \left(\epsilon_{A} \cdot kD \cdot e^{ M_A s_k } e^{ \epsilon_{A} D }  \right) \notag \\
                   & = \epsilon_{A} \cdot (k+1) D \cdot e^{ M_A \cdot (s_{k+1}) } e^{ \epsilon_{A} D },
        \end{align}
        which makes (\ref{tkResult}) legitimate for all $k$. Hence, applying any arbitrary matrix norm and substituting (\ref{eps}), (\ref{epsR}), and (\ref{tk+1Result}) in (\ref{D1t})  we get that 
    \begin{equation}\label{D1_Bound}
        |\Delta_1(t)| \leq  (l+1) \cdot \delta_1 |X(t)|,
    \end{equation}
    where \begin{equation}
        \delta_1(\epsilon) = \epsilon \cdot e^{M_A D}\left(M_K \cdot \left(\frac{D}{\tau_d} + 2\right) D \cdot e^{\epsilon D} + 1 \right),
    \end{equation}
    since
    \begin{equation}\label{switching bound}
        k \leq  \frac{D}{\tau_d} + 1. 
    \end{equation}   
    Next, we observe that for any variable $\Psi$ that can be $U,W,\Pi$
    \begin{align}\label{series_trans}
        \int_{t-D}^{t}{| \Psi(\theta) |^2 d \theta} &= \sum_{i=1}^{k+1}{\int_{t-D+s_{i-1}}^{t-D+s_i}{| \Psi(\theta) |^2 d \theta}}.
    \end{align}
    We can proceed now with bounding separately the terms in (\ref{d2t_new}), using the triangle inequality and the submultiplicative property. Applying any norm in (\ref{Z1}), along with (\ref{epsR}), (\ref{series_trans}) we get
        \begin{align}\label{Z1bnd}
           \sum_{n=1}^{k+1}  \left | \prod_{j=n}^{k}{e^{{A_i}(s_{j+1}-s_{j})}}  Z_{1,n}(t) \right| \leq e^{ M_A D} \cdot  \epsilon_{B} \cdot \int_{t-D}^{t}{|U(\theta)|d\theta}.
        \end{align}
        Similarly for the second term in (\ref{d2t_new}) we recall (\ref{TheorResult}), and hence,
        \begin{align}\label{Z2bnd}
            \sum_{n=1}^{k+1}  \left | \prod_{j=n}^{k}{e^{{A_i}(s_{j+1}-s_{j})}}  Z_{2,n}(t) \right| &\leq \epsilon_A \cdot D \cdot  e^{(M_A + \epsilon_A)\cdot D} M_B \notag \\
            & \quad \times \int_{t-D}^{t}{|U(\theta)|d\theta}.
        \end{align}
        Recalling (\ref{tk+1Result}), we get similarly
        \begin{align}\label{Z3bnd}
            \sum_{n=1}^{k+1} {\left| Z_{3,n}(t) \right|} &\leq \epsilon_A \cdot (k+1) \cdot D e^{(2 M_A + \epsilon_A)D} \cdot M_B \int_{t-D}^{t}{|U(\theta)|d\theta}.
        \end{align}
    Applying (\ref{eps}), (\ref{epsR}), (\ref{switching bound})--(\ref{Z2bnd}), and (\ref{Z3bnd}) in (\ref{d2t_new}) we get 
    \begin{equation}\label{D2_Bound}
       \left| \Delta_2(t) \right| \leq (l+1) \cdot \delta_2 \int_{t-D}^{t}{|U(\theta)|d\theta},
    \end{equation}
    where
    \begin{align}\label{deltab}
        \delta_2(\epsilon) &= \epsilon \cdot e^{M_A D} \left( M_K \left[e^{\epsilon D} \cdot D \cdot M_B \left(1+e^{M_A D}\left(\frac{D}{\tau_d}+1\right) \right)+1 \right] \right. \notag \\
        & \quad \left.+ M_B \right).
    \end{align}
    Applying (\ref{D1_Bound}), (\ref{D2_Bound}) in (\ref{Wt}) we arrive at (\ref{Wt_bound}) where
    \begin{align}
        \lambda(\epsilon) &= \max \{\delta_1(\epsilon),\delta_2(\epsilon)\}.
    \end{align}
\end{proof}
\end{lemma}

\begin{lemma}[{norm equivalency}]  \label{lemma_u(theta)}  
    For the inverse transformation (\ref{inverse_theta}), (\ref{2.19}) the following inequality holds for some positive constant $\nu_1$ 
    \begin{equation}\label{ut_bound}
       \int_{t-D}^{t}{| U(\theta) |^2 d \theta} \leq \nu_1 \left( |X(t)|^2 + \int_{t-D}^{t}{| W(\theta) |^2 d \theta} \right).
    \end{equation}
    Similarly, for the direct transformation (\ref{P_theta}), (\ref{W_theta}) it holds for some positive constant $\nu_2$
    \begin{equation}\label{wt_bound}
       \int_{t-D}^{t}{| W(\theta) |^2 d \theta} \leq \nu_2 \left( |X(t)|^2 + \int_{t-D}^{t}{| U(\theta) |^2 d \theta} \right).
    \end{equation}

\begin{proof}
    From (\ref{inverse_theta}) for the inverse transformation we apply Young's inequality to obtain 
    \begin{align}\label{U_theta_bound}
        \int_{t-D}^{t}{| U(\theta) |^2 d \theta} &\leq 2 \left( \int_{t-D}^{t}{ |W(\theta) |^2 d\theta} \right. \notag \\ 
        &\left. \quad + {M_K}^2  \int_{t-D}^{t}{  \left| { \Pi(\theta)   } \right| ^2 d \theta}  \right).
    \end{align}
    Setting $H_i=A_i + B_i K_i$, for $i=0,\ldots,l$, and using (\ref{2.5}), (\ref{2.19}) we obtain the following, in a similar manner of obtaining (\ref{P(t)}),
    \begin{align}\label{T_t}
        \int_{t-D}^{t}&{| \Pi(\theta) |^2 d \theta} = \sum_{i=1}^{k+1} \int_{t-D+s_{i-1}}^{t-D+s_i} \left| e^{H_{m_i}(\theta -t+D-s_{i-1})} \right. \notag \\ 
        & \times \left. \left( \prod_{n=1}^{i-1} e^{H_{m_n} (s_n - s_{n-1})} X(t) + \sum_{n=1}^{i-1}\prod_{j=n}^{i-2}{e^{{H_{m_{j+1}}}(s_{j+1}-s_{j})}} \right. \right.  \notag \\
        & \left. \left.  \times \int_{t-D+s_{n-1}}^{t-D+s_n}e^{H_{m_n}(t-D+s_n-\theta)}B_{m_n}W(\theta)d\theta \right) \right. \notag \\
        & \left. + \int_{t-D+s_{i-1}}^{\theta}e^{H_{m_i}(\theta - s)}B_{m_i}W(s)ds\right|^2 d\theta.    
    \end{align}
     Applying (\ref{series_trans}) and the triangle inequality in (\ref{T_t}), we reach (\ref{ut_bound}), where
    \begin{align}
        \nu_1 &= 2 \max \left\{ 2 M_{K}^2 D e^{2 M_H D}, 1+2 M_{K}^2 D^2 e^{2 M_H D} M_{B}^2 \right\}.
    \end{align}    
    Analogously, using the direct transformation from (\ref{P_theta}) and (\ref{W_theta}), we can similarly prove (\ref{wt_bound}) via (\ref{P(t)}), where
    \begin{align}
        \nu_2 &= 2 \max \left\{ 2 M_{K}^2 D e^{2 M_A D}, 1+2 M_{K}^2 D^2 e^{2 M_A D} M_{B}^2 \right\}.
    \end{align}\
\end{proof}
\end{lemma}

\begin{lemma}[{stability of target system}]\label{trans_stability}
    Let $P$ and $Q$ be as in (\ref{mean_delayf_stability}). For any $\epsilon < \epsilon^*$, where $\epsilon$ is defined in (\ref{eps}) and
    \begin{align}\label{eps_condition_2}
        \epsilon^* &= \min \left\{ \lambda^{-1}\left(\frac{1}{ \sqrt{2 e^D D \nu_1}}\right), \notag \right. \\
        &\quad \left. \lambda^{-1}\left(\frac{\lambda_{\min}({Q})}{ 2 M_B |P| \sqrt{2 e^D D \nu_1}} \right)  \right\},
    \end{align}  
    for $\lambda$ defined in Lemma \ref{lemma_w(t)}, there exist positive constants $\kappa$ and $\mu$ such that
    \begin{align}\label{trans_stability_result}
        \left| X(t) \right|^2 + \int_{t-D}^{t} W(\theta)^2 d\theta &\leq \kappa \left( |X(0)|^2 \right. \notag \\ 
        & \left. \quad  + \int_{-D}^{0} W(\theta)^2 d\theta \right) e^{-\mu t}, \quad t \geq 0.
    \end{align}

\begin{proof}
    For the target system we can now adopt the following common Lyapunov functional
    \begin{equation}\label{LyapunovFnc}
        V(t) = X(t)^T P X(t) + b \int_{t-D}^{t} e^{(\theta+D-t)}W(\theta)^2 d\theta.
    \end{equation}
    
    Calculating the derivative of (\ref{LyapunovFnc}), along the solutions of the target system we obtain
    \begin{align}
        \dot{V}(t) &\leq -X(t)^T  Q  X(t) + B_{\sigma(t)}^T W(t-D) \notag \\ &\qquad \times P X(t) + X(t)^T P B_{\sigma(t)} W(t-D) + b e^D W(t)^2 \notag \\
        & \quad - b W(t-D)^2 - b\int_{t-D}^{t} e^{(\theta+D-t)}W(\theta)^2 d\theta. \label{dv_t_1}
    \end{align}
    Observing $-X(t)^T Q X(t) \leq - \lambda_{\min}({Q}) |X(t)|^2$ and 
    \begin{align}\label{BTWT}
        B_{\sigma(t)}^T W(t-D)^T P X(t) + X(t)^T &P B_{\sigma(t)} W(t-D) \leq \notag \\
        & 2 \left|X(t)^T P M_B W(t-D)\right|,
    \end{align}
    applying Young's inequality, and choosing 
    \begin{equation}
        b = \frac{ 2 (M_B|P|)^2 }{ \lambda_{\min}({Q}) }, 
    \end{equation}
    then we get from (\ref{dv_t_1}) that
    \begin{align}
    \label{dv_t_11}
        \dot{V}(t) &\leq -\frac{1}{2} \lambda_{\min}({Q}) |X(t)|^2 \notag \\
        & \quad + be^D W(t)^2 - b\int_{t-D}^{t} e^{(\theta+D-t)}W(\theta)^2 d\theta.
    \end{align}
    Using Lemmas \ref{lemma_w(t)}  and \ref{lemma_u(theta)} we get 
    \begin{align}\label{wt_sqrt11}
        W(t)^2 &\leq 2  \lambda^2(\epsilon)(D \nu_1 + 1) |X(t)|^2 + 2 \lambda^2(\epsilon) D \nu_1 \notag \\
        &\qquad \times \int_{t-D}^{t}e^{(\theta+D-t)}{W(\theta)^2 d\theta}.
    \end{align}    
    Employing (\ref{wt_sqrt11}) in (\ref{dv_t_11})  we get
    \begin{align}\label{dv_t_12}
        \dot{V}(t) &\leq - \left( \frac{1}{2} \lambda_{\min}({Q}) - 2 b e^D \lambda^2(\epsilon)(D \nu_1 + 1) \right)|X(t)|^2 \notag \\
                   &\quad -b \int_{t-D}^{t}{ \left[1 - 2 e^D  \lambda^2(\epsilon) D \nu_1         \right] e^{(\theta+D-t)} W(\theta)^2 d\theta } . 
    \end{align}
    In order to preserve negativity in (\ref{dv_t_12}) it is required that
    \begin{align}\label{d_v_t13}
        \frac{1}{2} \lambda_{\min}({Q}) - 2 b e^D \lambda^2(\epsilon)(D \nu_1 + 1) &> 0, \notag \\
        1 - 2 e^D  \lambda^2(\epsilon) D \nu_1 &> 0 ,
    \end{align}
    which hold under the restriction on $\epsilon$ in the statement of the lemma.
    From (\ref{d_v_t13}) and the comparison principle it follows that
    \begin{equation}
        V(t) \leq e^{-\mu t}V(0),\quad t \geq 0,
    \end{equation}
    where 
    \begin{align}
        \mu &= \min \left \{ 1-2e^D \lambda^2(\epsilon) D \nu_1, \right. \notag \\
        &\left. \quad \frac{  \frac{1}{2}\lambda_{\min}(Q) - 2 be^D \lambda^2(\epsilon)(D \nu_1 + 1) }{\lambda_{\max}(P)} \right \}.
    \end{align}
    Thus, we arrive at (\ref{trans_stability_result}) with
    \begin{equation}
        \kappa = \frac{\max \left\{ \lambda_{\max}(P), \frac{2 M_{B}^2 |P|^2}{\lambda_{\min}(Q) } e^D \right\}}{\min \left\{ \lambda_{\min}(P), \frac{2 M_{B}^2 |P|^2}{ \lambda_{\min}(Q)  } \right\}}.
    \end{equation}
\end{proof}

\begin{customproof}[{Proof of Theorem \ref{Exponential Stability}}: ]
    Combining (\ref{ut_bound}), (\ref{wt_bound}), and (\ref{trans_stability_result}) we get (\ref{stability equation}) where    
    \begin{align}
    \rho &= \sqrt{\frac{2\nu_1\nu_2}{\kappa}}, \quad 
    \xi  = \frac{\mu}{2}. 
    \end{align}
\end{customproof}
\end{lemma}

\section{Simulation Results}\label{sec4}
\begin{figure}[ht]
    \centering
    \includegraphics[width=8.5 cm]{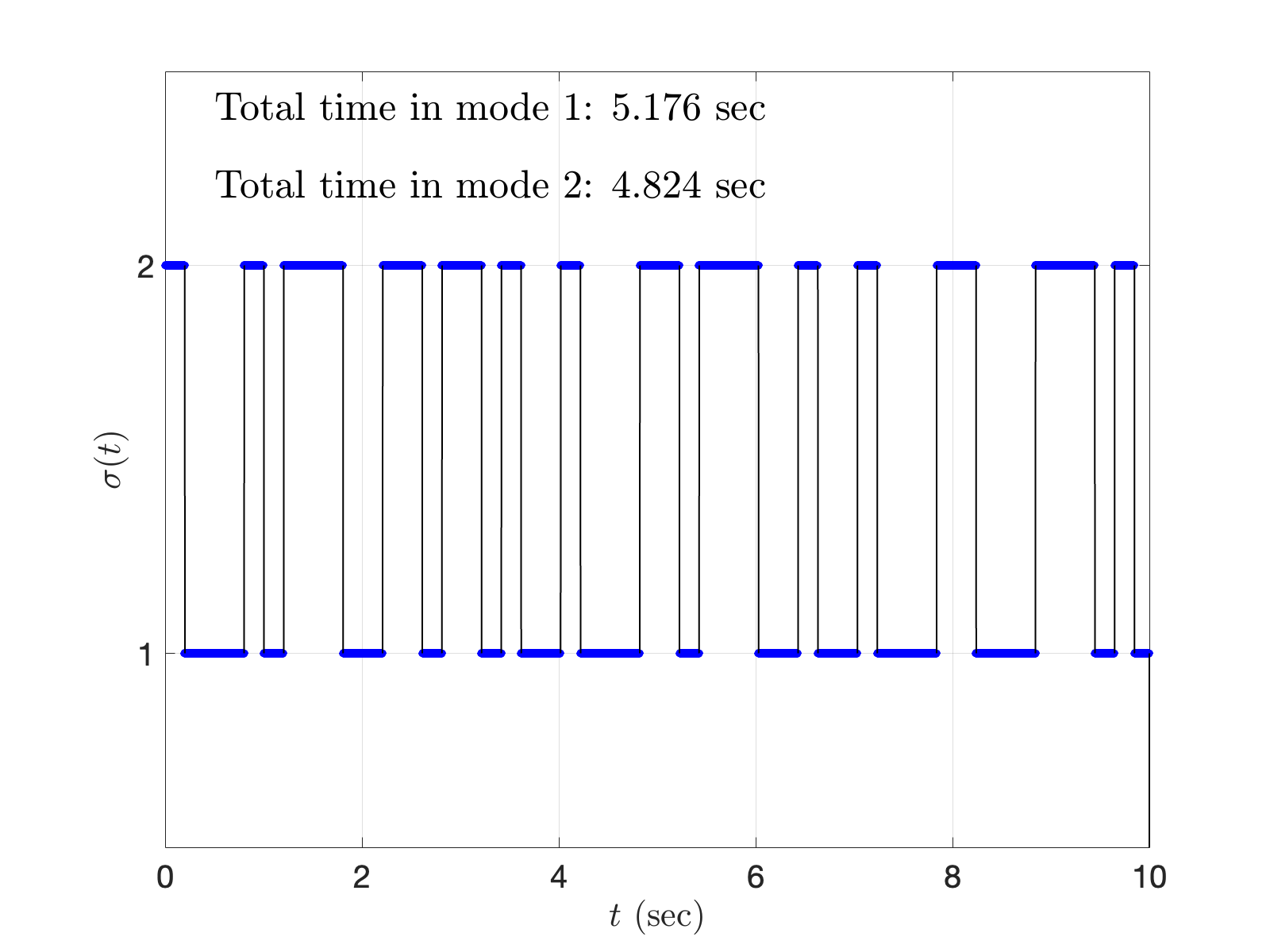}
    \caption{Evolution of switching signal $\sigma(t)$ for all the case studies.
    \label{fig3}}
\end{figure}

\begin{figure}[ht]
    \centering
    \includegraphics[width=8.5 cm]{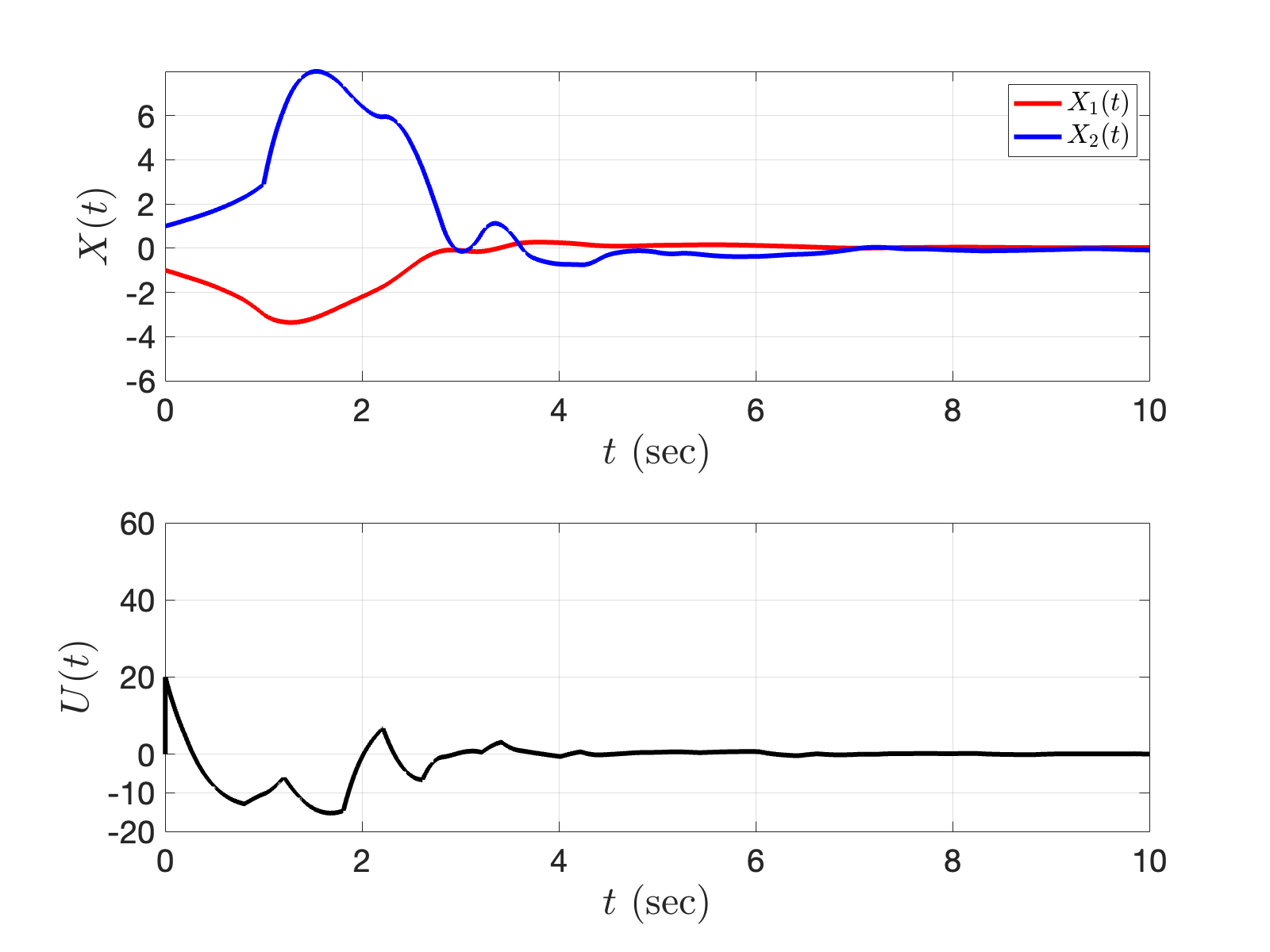}
    \caption{Evolution of state $X(t)$ and control input $U(t)$ for system (\ref{1.1}) with (\ref{ex1}), (\ref{ex1_dyn}), under controller (\ref{1.5}) with (\ref{ex1_k}). 
    \label{fig2}}
    \end{figure}

Consider the switched system (\ref{1.1}) with the unstable subsystems \( A_1 \) and \( A_2 \) as
\begin{equation}\label{ex1}
    A_1 = \begin{bmatrix}
    2 & 1 \\
    0 & 1
\end{bmatrix}, \quad
A_2 = \begin{bmatrix}
    2.3 & 1.1 \\
    0.05 & 1.2
\end{bmatrix},
\end{equation}
and input matrices \( B_1 \) and \( B_2 \) as
\begin{equation}\label{ex1_dyn}
    B_1 = \begin{bmatrix}
    0 \\
    1
\end{bmatrix}, \quad
B_2 = \begin{bmatrix}
    0 \\
    1.02
\end{bmatrix}.
\end{equation}
Next we choose 
\begin{equation}\label{ex1_k}
K_1 = \begin{bmatrix} -12 & -6 \end{bmatrix}, \quad K_2 = \begin{bmatrix} -12.6961 & -6.3725 \end{bmatrix},
\end{equation}
which places the closed-loop poles at $-1,-2$ for each $i=1,2$. We set $D=1$, $\tau_d=0.2$, and initial conditions $X_0 = \begin{bmatrix} -1 & 1 \end{bmatrix}^{T}$, $U(s)=0$, for $s \in [-D,0)$. Fig. \ref{fig3} shows the evolution of the switching signal over the simulation time. Active mode is indicated by the blue lines and in black the switching instants. We obtain for these parameters $\epsilon^*=7 \times 10^{-5}$ approximately and $\epsilon=0.7895$. The system's response under (\ref{1.5}) is depicted in Fig. \ref{fig2}. This result is intended to show that $\epsilon^*$ may be highly conservative. 

\begin{figure}[t]
    \centering
    \includegraphics[width=8.5 cm]{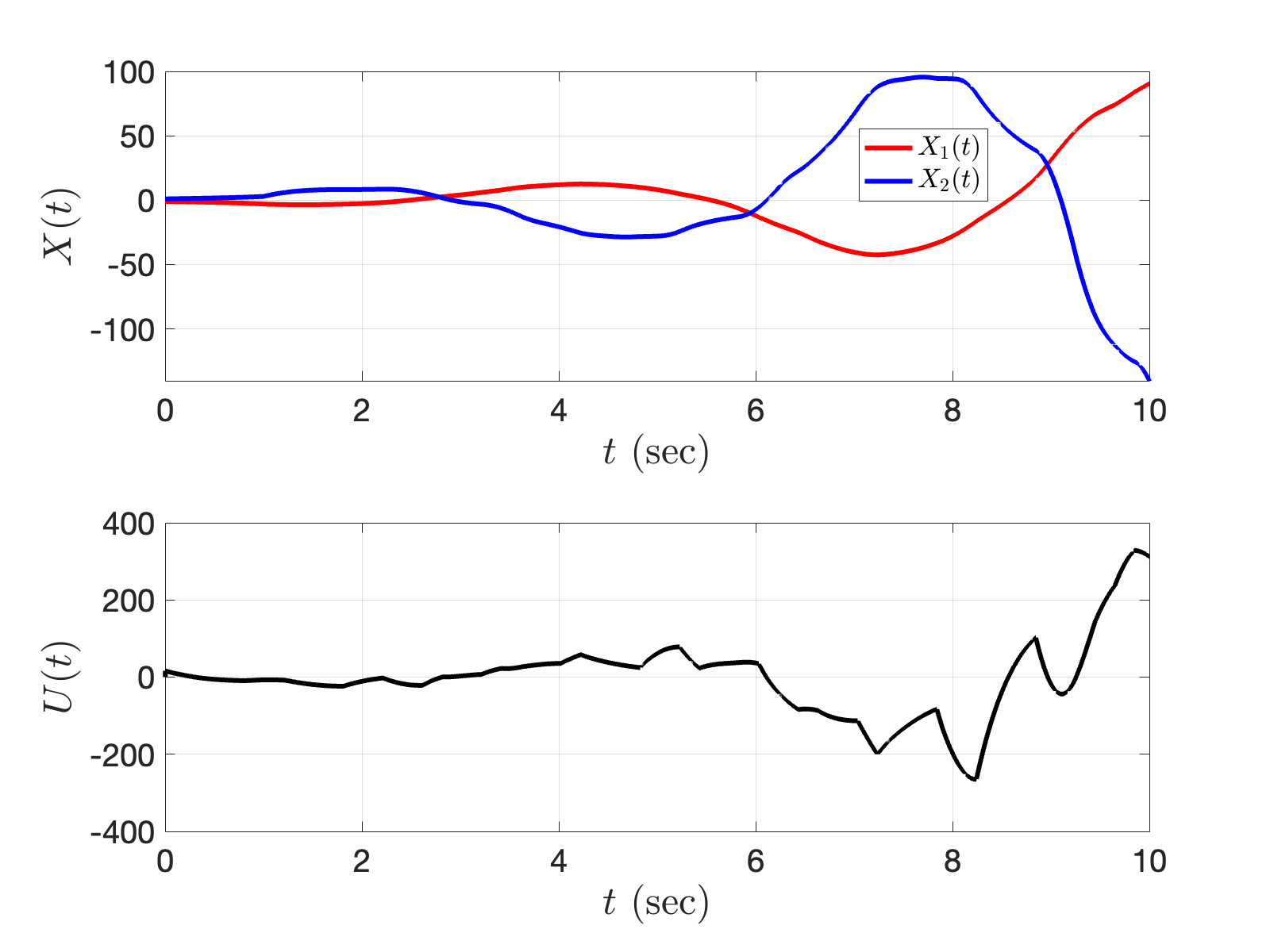}
    \caption{Evolution of state $X(t)$ and control input $U(t)$ for system (\ref{1.1}) with (\ref{ex1}), (\ref{ex1_dyn}), under controller (\ref{onecontrol}) for $i=1$ with $K_1$ as in (\ref{ex1_k}). 
    \label{fig4}}
    \end{figure}
\begin{figure}[t]
    \centering
    \includegraphics[width=8.5 cm]{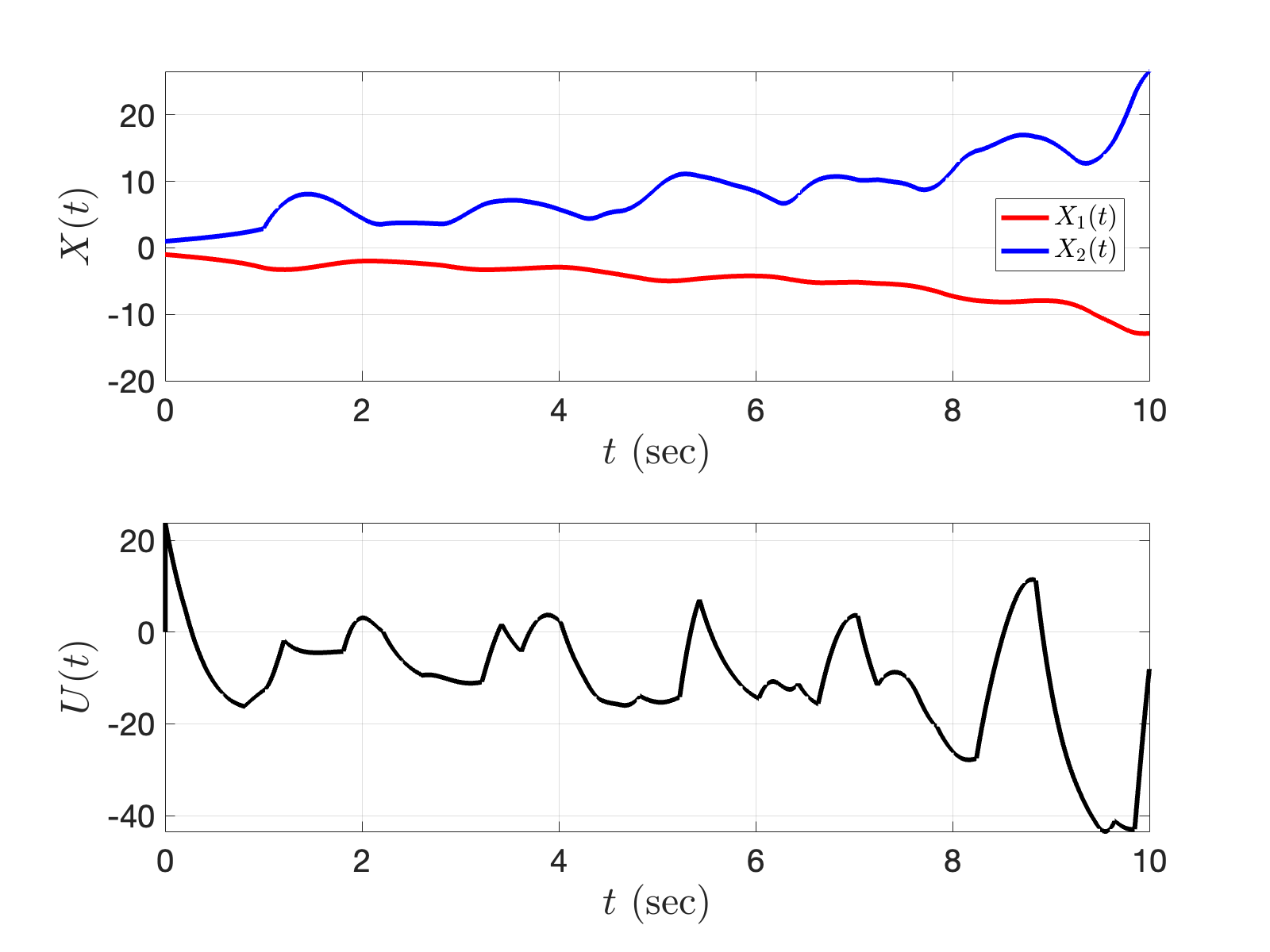}
    \caption{Evolution of state $X(t)$ and control input $U(t)$ for system (\ref{1.1}) with (\ref{ex1}), (\ref{ex1_dyn}), under controller (\ref{onecontrol}) for $i=2$ with $K_2$ as in (\ref{ex1_k}). 
    \label{fig5}}
    \end{figure}  
\begin{figure}[ht!]
    \centering
    \includegraphics[width=8.5 cm]{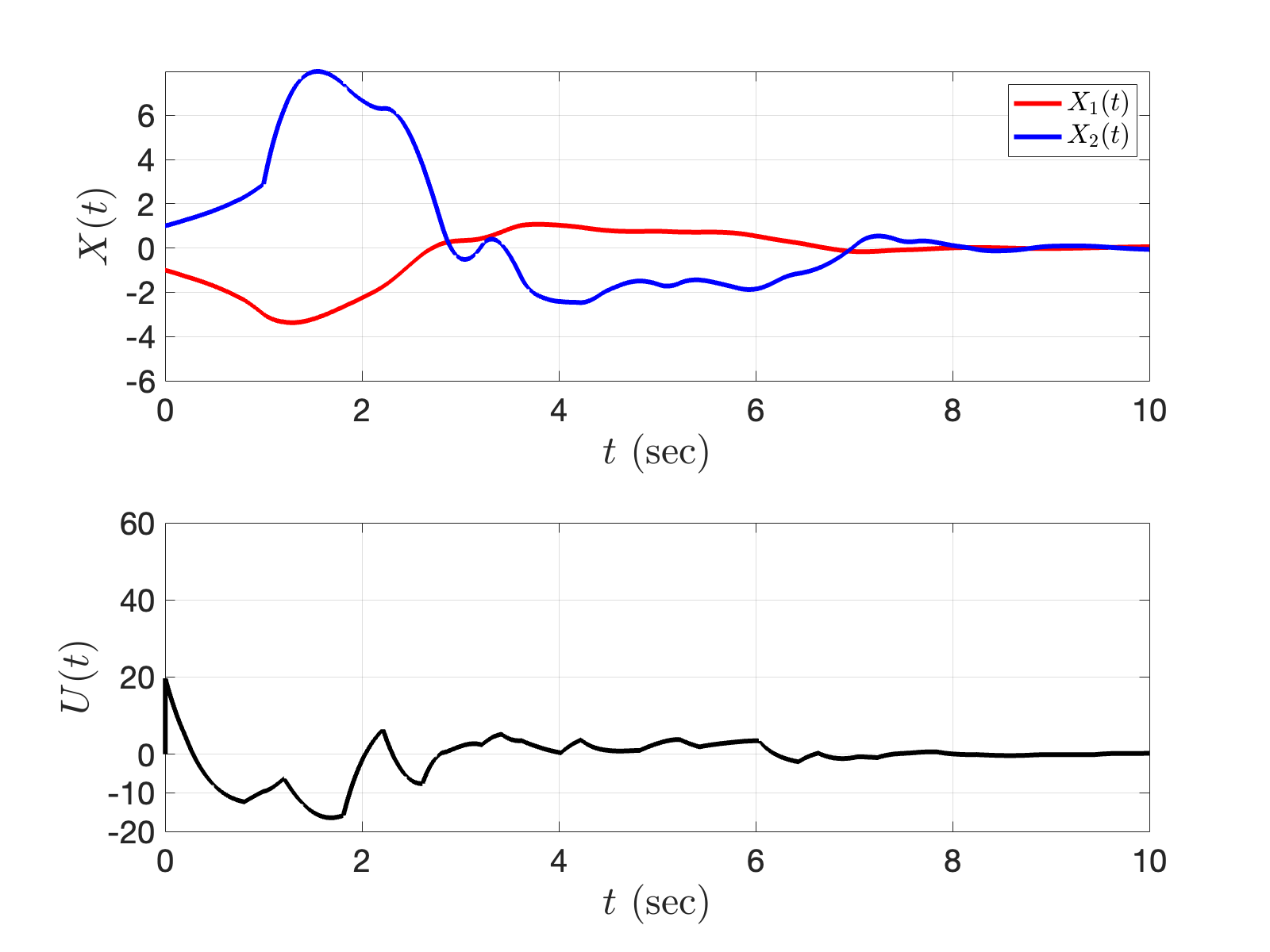}
    \caption{Evolution of state $X(t)$ and control input $U(t)$ for system (\ref{1.1}) with (\ref{ex1}), (\ref{ex1_dyn}), under controller (\ref{Uavg}) with (\ref{ex_k_avg}). 
    \label{fig6}}
    \end{figure}
    
Next, we compare the same system's behaviour operating under two different controllers.  The first controller, motivated by the fact that the subsystems are close,  assumes that the system is operating only in one mode as 
\begin{equation}\label{onecontrol}
U(t) = K_i \left(e^{{A_i}D} X(t) + \int_{t-D}^{t} {e^{{A_i}(t-\theta)}B_iU(\theta) \, d\theta}\right), \quad i=1,2.
\end{equation} 
We choose the same $K_i$ gains to retain the same closed-loop poles, as in (\ref{ex1_k}), for a fair comparison. Fig. \ref{fig4} shows the response of the system under the controller (\ref{onecontrol}) for $i=1$ and Fig. \ref{fig5} shows the response under the same controller for $i=2$. It is evident that, in both cases, the performance of the respective closed-loop systems is not the desired one. This is attributed to controller (\ref{1.5}) achieving a more accurate prediction of the future values of the state. 

Secondly, we compare the controller's performance, as shown in  Fig. \ref{fig2}, with the controller from \cite{my_ACC} defined as
\begin{equation}\label{Uavg}
{U}(t) = \bar{K} \left(e^{\bar{A}D} X(t) + \int_{t-D}^{t} {e^{\bar{A}(t-\theta)}\bar{B}U(\theta) \, d\theta}\right),
\end{equation}
where $\bar{A},\bar{B}$ are chosen as the mean matrices of the sets $\{A_1,A_2\}$ and $\{B_1,B_2\}$, respectively, where as mean matrix we define the element-wise mean, and $\bar{K}$ is selected as
\begin{equation}\label{ex_k_avg}
   \bar{K}= \begin{bmatrix}
    -12.3515 & -6.1881
\end{bmatrix},
\end{equation}
to place the closed-loop poles of the expected system (corresponding to matrix $\bar{A} + \bar{B} \bar{K}$), for the purposes of the comparison, at $-1,-2$. The system's response under (\ref{Uavg}) is depicted in Fig. \ref{fig6}. Controller (\ref{Uavg}) can stabilize the system effectively. On the other hand, convergence is slower as compared with Fig. \ref{fig2}, while better transient response is also achieved with controller (\ref{1.5}). In general, we observed in our simulation studies that the control design developed here behaves more efficiently when dwell time decreases, which may be explained as (\ref{1.5}) providing, in such cases, a predictor-based law that is closer to the exact predictor-feedback law for the switched system (i.e., a controller that employs (\ref{P(t)})), since it involves an average of all possible (for each subsystem) exact predictor-feedback laws. On the other hand, we observed that (\ref{Uavg}) may be more robust when epsilon increases, i.e., when the differences among system's matrices (and controller's gains) increase. However, further investigations are needed to make concrete conclusions about the advantages/disadvantages of (\ref{Uavg}) as compared with (\ref{1.5}).

\section{Conclusions and Discussion}\label{sec5} 
We proposed a  predictor-based control law for switched linear systems, subject to arbitrary switching and input delay, where the future switching signal is unknown. To overcome these challenges, we developed a control law based on averaging exact predictor-feedbacks. Uniform stability is proved, by necessarily restricting the differences between system's matrices and controller's gains, without however imposing constraints on delay and dwell time values. Numerical simulations illustrated the efficiency of the control strategy, also in comparison with alternative, predictor-based control laws. We are currently investigating the advantages/disadvantages of control laws (\ref{1.5}) and (\ref{Uavg}), both theoretically and in simulation. As future work, we aim at developing new predictor-based control laws alleviating the constraints on system's matrices and controller's gains, by taking into account possible knowledge of a dwell time, for constructing more accurate predictors of the future state. It is anticipated that this requires imposing certain constraints on the value of dwell time.
\appendix{}
\section{}

\renewcommand{\thelemmaA}{A.\arabic{lemmaA}}
\renewcommand{\thepropositionA}{A.\arabic{propositionA}}
\renewcommand{\theequation}{A.\arabic{equation}}

\setcounter{equation}{0}

\begin{propositionA}\label{prop1}
    For any two $n \times n$ matrices $Y_1$, $Y_2$ the following inequality holds:
    \begin{align}
       \left| e^{Y_1+Y_2}-e^{Y_1} \right| \leq |Y_1| e^{ |Y_1| }e^{ |Y_2| },
    \end{align} 
    where $|\cdot|$ denotes an arbitrary matrix norm.

\begin{proof}
Using the series for the matrix exponential we get
\begin{align}
    e^{Y_1+Y_2}-e^{Y_1} &= \sum_{k=0}^\infty\frac1{k!}\,[(Y_1+Y_2)^k-Y_1^k] \notag \\ 
    &= \sum_{k=1}^\infty\frac1{k!}\;
    \sum_{(j_1,\dots,j_k)\in\{1,2\}^k\setminus\{(1,\dots,1)\}}\;
    \prod_{i=1}^k Y_{j_i}. 
\end{align}
Considering that $|AB| \leq |A||B|$ and using triangle inequality 
\begin{align}
    \left| e^{Y_1+Y_2}-e^{Y_1} \right| 
    &\le \sum_{k=1}^\infty\frac1{k!}\;
    \sum_{(j_1,\dots,j_k)\in\{1,2\}^k\setminus\{(1,\dots,1)\}}\;
    \prod_{i=1}^k |Y_{j_i}| \notag \\ 
    &= \sum_{k=1}^\infty\frac1{k!}\;
    \sum_{m=1}^k \binom km |Y_1|^{k-m} |Y_2|^{m} \notag \\  
    &= \sum_{k=1}^\infty\frac1{k!}\;[(|Y_1|+|Y_2|)^k-|Y_1|^k] \notag \\ 
    &= e^{|Y_1|+|Y_2|}-e^{|Y_1|} \notag \\ 
    &= e^{|Y_1|+|Y_2|}(1-e^{-|Y_2|}) \notag \\
    &\le |Y_2|e^{|Y_1|+|Y_2|}.
\end{align}
\end{proof}
\end{propositionA}
\stepcounter{propositionA}
\bibliographystyle{amsplain}

\end{document}